\documentclass[letterpaper,jpc,twocolumn,floatfix]{revtex4}  %two column format

\usepackage[normalem]{ulem}  %for underlines and strike throughs 
\usepackage{amsmath}
\usepackage[mathcal]{euscript}
\usepackage{mathrsfs}
\usepackage{float}
\usepackage[usenames]{color} %for colored text
\usepackage{graphicx}     %use this if you're using pdflatex to include hyperlinks
\usepackage{hyperref}           %include this if you're using hyperlinks

\begin{document}

\title{Analysing kinetic transition networks for rare events}
\date{\today}
\author{Jacob D. Stevenson}
\author{David J. Wales}
\affiliation{University Chemical Laboratories, University of Cambridge, Lensfield Road, Cambridge CB2 1EW, UK}

\begin{abstract}

  The graph transformation approach is a recently proposed method for computing mean first passage times,
  rates, and committor probabilities for kinetic transition networks. 
  Here we compare the performance to existing linear algebra methods, focusing on
  large, sparse networks. 
  We show that graph transformation provides a much more robust framework,
  succeeding when numerical precision issues
  cause the other methods to fail completely.  These are precisely the situations that correspond
  to rare event dynamics for which the graph transformation was introduced.

\end{abstract}

\maketitle

%\tableofcontents
%\bibliographystyle{naturemagurl}
\bibliographystyle{aip}
%\bibliographystyle{/people/jake/research/apsrev}
%\bibliographystyle{apsrev}

%\section{Introduction}

The kinetics of many complex physical processes can be described by kinetic transition
networks \cite{NoeF08,pradag09}. % \cite{norris.1997,NoeF08,pradag09}.
In these networks the discrete states correspond to the nodes of a graph, whose edges
encode the underlying transitions. % that are allowed by the underlying physical process.
In many situations the Markov approximation holds and transitions between the states
are taken to be independent random processes.
% A rate constant assigned to each directed edge describes the expected waiting time for that transition to occur.
These kinetic transition networks can also be viewed as continuous time Markov processes.
They are widely used in the physical sciences, and also in 
other fields such as finance \cite{masoliver.2006}
and modelling of social networks \cite{acemoglu.2013}.
In protein folding studies the states and rates are often defined by data gathered from molecular dynamics
simulations \cite{NoeSVRW09}.  Alternatively, the states may be local minima on 
the potential energy landscape, where the rate constants are
calculated from unimolecular rate theory \cite{wales02,wales04}.

Rate constants are local properties specifying the time scale on which
direct transitions between states occur.  However, we usually want to calculate
experimental observables, such as the mean first passage time (MFPT) between two states.
These global properties of the network can be computed stochastically, e.g.~using kinetic Monte Carlo simulations
\cite{HenkelmanJ01}, or, 
if the number of states is small enough, by
directly solving the master equation through matrix diagonalization.
Unfortunately, stochastic methods are approximate and can be rather slow to converge, while
the exact methods tend to suffer from numerical precision problems \cite{glowacki.2012} due to poorly conditioned matrices.  
This situation is likely to be encountered for rare events, where the range of relaxation times can 
span many orders of magnitude.
Here we discuss the performance of a recently introduced method for
computing global kinetic quantities called the new graph transformation (NGT) approach \cite{Wales09}
and compare it to existing methods.
We show how NGT overcomes the
numerical precision problems that plague other methods with little additional
overhead in terms of computing time.

%\section{Methods}

Consider a kinetic transition network \cite{NoeF08,pradag09,Wales10a} with 
$N$ nodes and $E$ edges.  To each
edge $u \to v$ is associated a rate constant $k_{uv}$.  
It is convenient to define the rate matrix $R_{uv}$ as
\begin{equation}
    R_{uv} = k_{uv}  \text{ \quad for } u \ne v \quad {\rm with} \quad
    \sum_v R_{uv} = 0.
\end{equation}
The second condition specifies that the diagonal components are given by $R_{uu} = - \sum_v k_{uv}$.
The kinetic transition network can be equivalently
expressed in terms of transition probabilities $P_{uv}$ and waiting times $\tau_u$,
where 
\begin{equation}
  \tau_u = \left( \sum_v k_{uv} \right)^{-1}
  \text{ \quad and \quad }
  P_{uv} = \tau_u  k_{uv}.
\end{equation}
We further specify a product group $A$ and a reactant group $B$, which may consist of multiple nodes,
for which we want to compute rates and MFPTs.

We can specify the MFPT $T_{uB}$, the mean time for a trajectory starting at $u$ to reach a node in $B$,
in terms of the MFPTs of the neighbours of $u$ as 
$T_{uB} = \tau_u + \sum_x P_{ux} T_{xB}$.
Written in terms of $R$ this becomes \cite{norris.1997}
\begin{equation}
    \sum_{x \notin B} R_{ux} T_{xB} = -1    \text{\quad \quad for } u \notin B
    .
\end{equation}
This is a system of linear equations, which can be solved for the vector
$\{T_{uB} | u \notin B\}$.  
If the product group $A$
contains more than one node,
the transition rate from $A$ to $B$ is an average over
the inverse MFPT for each
element $a$ in $A$, weighted by their equilibrium occupation probability, $p^{\text{eq}}_a$
\begin{equation}
  k_{AB} = \left< \frac{1}{ T_{aB} } \right>_{a \in A} = 
  \frac{1}{\sum_{a \in A} p^{\text{eq}}_a}  \sum_{a \in A} \frac{p^{\text{eq}}_a}{T_{aB}}
  .
  \label{eqn:rate}
\end{equation}

Committor probabilities, the probability that node $u$ will reach $B$ before it reaches $A$, are defined such that 
$q_u = 0$ if $u \in A$, and $q_u = 1$ if $u \in B$.
For $u \notin A \cup B$ they can be found via the relation
$q_u = \sum_{v} P_{uv} q_v$.
In terms of R, this becomes
\begin{equation}
    \sum_{x \notin A \cup B} R_{ux} q_x = - \sum_{b\in B} R_{ub}  \text{ \quad for } u \notin A \cup B
    .
\end{equation}
which can be solved numerically for the vector $\{q_u | u \notin A \cup B\}$ \cite{norris.1997,metznersv09}.

%\subsection{NGT: mean first passage time}

The NGT method is a deterministic graph renormalization 
procedure \cite{Wales09,trygubenkoW06a,trygubenkow06b} to compute the exact MFPT
averaged over the product group $B$, for any member of the
reactant group $A$.
We use `renormalization' in the sense of real space renormalization group theory \cite{landau.2005}.
Nodes are deterministically removed and the waiting times and branching probabilities of
neighbouring nodes are updated so that the MFPT for any reactant state averaged
over all the product states is preserved; the proof does not require
detailed balance \cite{Wales09,trygubenkoW06a,trygubenkow06b}.
% An important
% feature of this algorithm is that 
Each node $u$ is also assigned a loop edge
$u \to u$ pointing back to itself.  In the typical case, the self-transition
probabilities $P_{uu}$ will all be zero initially, but will take non-zero
values after renormalization.  The transition probabilities always satisfy
$\sum_v P_{uv} = 1$.

Upon removing node $x$,
the updated transition probabilities are found by summing the
direct path from $u$ to $v$ and all paths through $x$
\begin{align}
  P_{uv}  \to &  P_{uv} + P_{ux} P_{xv} \sum_{m=0}^{\infty} P_{xx}^m \nonumber \\
  \to & P_{uv} + \frac{ P_{ux} P_{xv} } { 1 - P_{xx} }
  .
  \label{eqn:P_update}
\end{align}
The self-transition probabilities $P_{uu}$ are updated according to the
same equation.
Similarly, the updated waiting time $\tau_u$ is found by computing the mean time
to reach one of the neighbours of $u$ (excluding $x$), or to return to $u$.
\begin{align}
  \tau_u  \to &  \sum_{v \ne x} \left\{ 
P_{uv} \tau_u + P_{ux} P_{xv}
\sum_{m=0}^{\infty} \left[ \tau_u + (m+1) \tau_x \right] P_{xx}^{m} 
\right\} \nonumber \\
  \to &  \tau_u + \frac{P_{ux}  \tau_x} {1 - P_{xx}}
  .
  \label{eqn:tau_update}
\end{align}
Equations \ref{eqn:P_update} and \ref{eqn:tau_update} constitute the NGT graph renormalization procedure.

We wish to compute the mean first passage time from node $a \in A$ to the
product group of nodes $B$.
%, however,
%the full algorithm is most easily described if we first assume that $B$ 
%contains only one node $b$.  Nodes are
%iteratively removed, (updating the graph attributes according to \ref{eqn:P_update} and \ref{eqn:tau_update}) until the only two %remaining nodes are $a$ and
%$b$.  
%At this point, the renormalized probability
%$P_{ab}$ is simply the probability that a trajectory starting at $a$
%will end up at $b$ before returning to $a$.  Similarly, the mean first passage
%time from $a$ to $b$ is $\tau_a / P_{ab}$.  
%The transition rate from $a$ to $b$ is simply the inverse of the
%mean first passage time.  The rates and probabilities from $b \to a$ are read
%from the renormalized graph in the same way.  
%If there is more than one element in $B$ the calculation of rates from $a \to B$
%is nearly as straightforward.  
Nodes are
iteratively removed from the graph, updating the graph attributes according to \ref{eqn:P_update} and \ref{eqn:tau_update}),
until the only remaining nodes are $a$ and $b\in B$.
In this reduced graph, every trajectory starting from $a$ (except those following the loop edge $a \to a$) will transition directly
to $B$.
The MFPT is, then, found from the relation $T_{aB} = \tau_a + P_{aa} T_{aB}$, which leads directly to
\begin{equation}
  T_{aB} = \frac{ \tau_a }{ 1 - P_{aa} }
  .
  \label{eqn:mfpt}
\end{equation}
The rate from $A$ to $B$ can then be computed via equation \ref{eqn:rate}.
The calculation, in practice, is performed in two phases.  First, the intervening
nodes (not in $A$ or in $B$) are all removed from the graph.  
% In the second phase we first make a backup copy of the graph.  
Then for each node 
$a\in A$ we compute $T_{aB}$ by removing from the graph all nodes in $A$ except $a$.
The rates $B \to A$ can be computed in a similar manner.

Solving for the committor probability is straightforward in the NGT framework.
Using equations \ref{eqn:P_update} and \ref{eqn:tau_update}
we first remove all nodes in the graph except those in $A$, $B$, and $u$ itself.  We can then
read off the committor probability as \cite{Wales09}
\begin{equation}
  q_u =
   \frac{ \sum_{b \in B} P_{ub} }{ \sum_{x\in A \cup B} P_{ux}}
   %= \frac{ \sum_{b \in B} P_{ub} }{ 1-P_{uu}}  
   \text{ \quad for } u \notin A \cup B
   .
  \label{eqn:ngt_committors}
\end{equation}
%where the second equality only holds if $u$ is not in $A \cup B$.
The denominator can also be written as $1-P_{uu}$.
In reference \cite{Wales09} the committor is defined as $C_u^B$ in a slightly different way.  
For $u \notin A \cup B$ it is equivalent
to $q_u$, but $C_u^B$ can take non-zero values for $u \in A$.

In equations 
\ref{eqn:P_update}, 
\ref{eqn:tau_update}, 
and
\ref{eqn:mfpt}
we can compute $1-P_{uu}$ in two different ways via the relation 
\begin{equation}
  1 - P_{uu} = \sum_{v \ne u} P_{uv} .
\end{equation}
This procedure allows us to maintain numerical precision when $P_{uu}$ is very small
and when $P_{uu}$ is very close to 1.

There are some important differences between NGT and the linear algebra method.  Solving the system
of linear equations results in the MFPT (or committor) for every node in the graph.  In contrast, the MFPTs (and committors) 
for NGT are treated one node at a time.  If we 
are interested in $k_{AB}$, and $A$ is not too large, the additional overhead is minor.
On the other hand, when computing $k_{AB}$ with the NGT method, the reverse rate $k_{BA}$ is obtained essentially for free.

%The steady state rate from $A$ to $B$ can be computed straightforwardly from the committors via the relationship
%\begin{equation}
%k^{SS}_{AB} = \left< \sum_v k_{av} q_v \right>_{a \in A}
%.
%\end{equation}
%The average is over the nodes in $A$ weighted by their equilibrium occupation probability, as in equation \ref{eqn:rate}.
%In the NGT, a more convenient formulation is used.  Once all the intermediate nodes have been removed, the steady state rate
%is written
%\begin{equation}
%k^{SS}_{AB} = \left< \frac{ \sum_{b \in B} P_{ab}}{ \tau_a^{I} } \right>_{a \in A}
%,
%\end{equation}
%where $\tau_a^{I}$ refers to the waiting time in the initial graph, before any renormalization \cite{Wales09}.

\begin{figure}
  	\includegraphics[width=\linewidth]{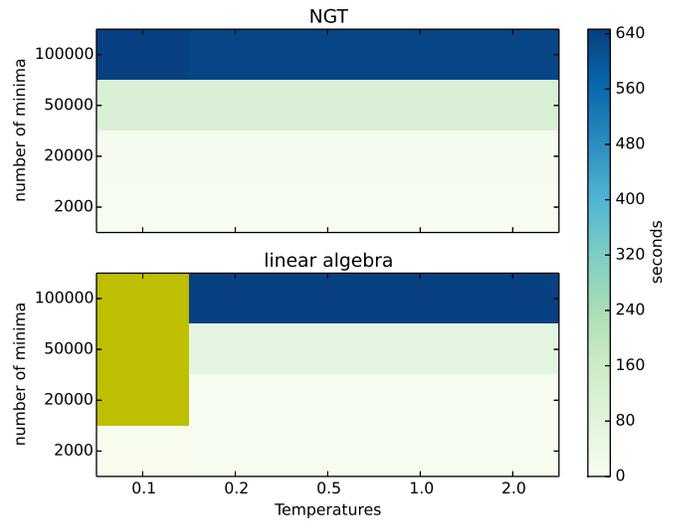}
  \caption{CPU time required to compute rates with NGT and a sparse linear
  algebra solver for the LJ$_{38}$ cluster as a function of temperature and
  the number of nodes.  Yellow indicates that the method failed.}
  \label{fig:lj38_times}
\end{figure}

\begin{figure}
  	\includegraphics[width=\linewidth]{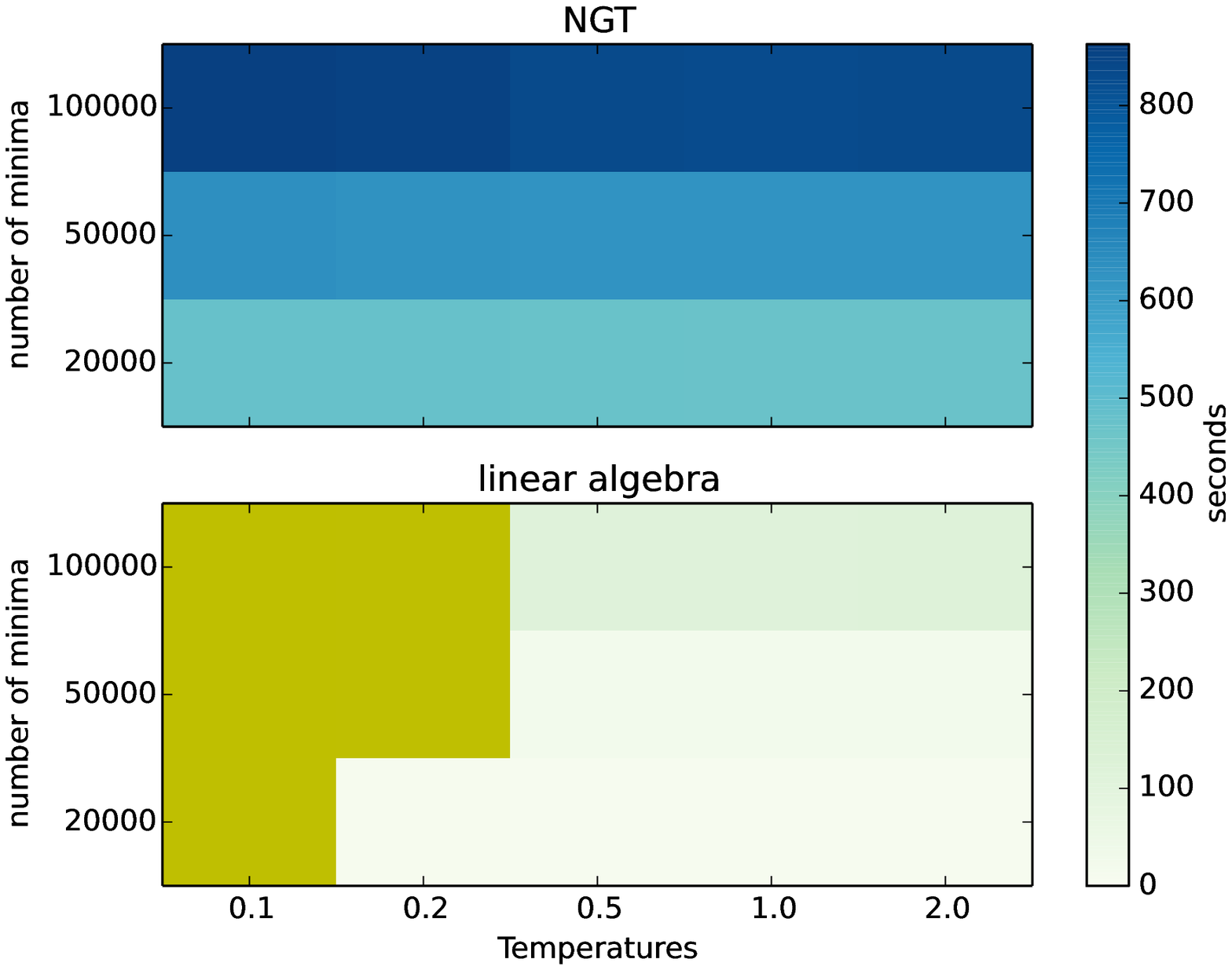}
  \caption{CPU time required to compute rates with NGT and a sparse linear
  algebra solver for the LJ$_{75}$ cluster as a function of temperature and
  the number of nodes.  Yellow indicates that the method failed.}
  \label{fig:lj75_times}
\end{figure}

\begin{figure}
  	\includegraphics[width=\linewidth]{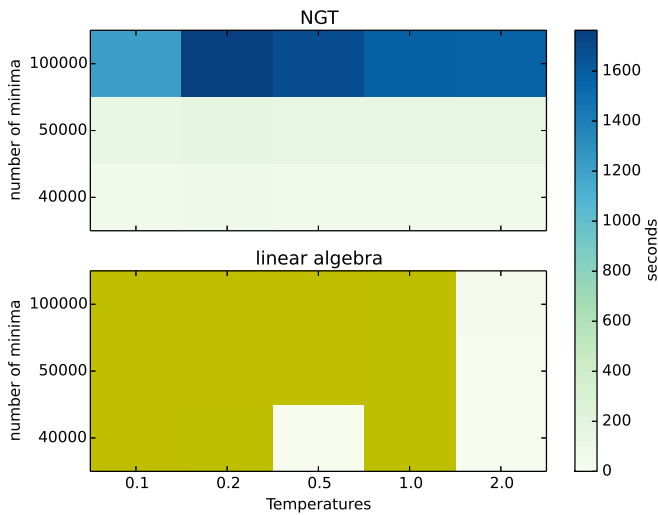}
  \caption{CPU time required to compute rates with NGT and a sparse linear
    algebra solver for the three-stranded $\beta$-sheet peptide Beta3s as a
    function of temperature and the number of nodes.  Yellow indicates that the method
  failed.}
  \label{fig:beta3s_times}
\end{figure}

%\section{Discussion}

To compare the performance of the NGT method with the linear algebra approach
we chose several benchmark systems that are representative of important
problems in rare event dynamics.  We consider two
Lennard-Jones clusters of 38 atoms \cite{doyemw99} and 75 atoms \cite{doyemw99}, denoted LJ$_{38}$
and LJ$_{75}$, along with the three-stranded
$\beta$-sheet peptide Beta3s \cite{carrw08}.  The networks were
generated in previous discrete path sampling studies \cite{wales02,wales04}.  The nodes of these networks represent minima on
the potential energy landscape (locally stable configurations), while the edges correspond to
transition states connecting the minima.  These stationary points
were computed numerically using geometry optimization
techniques \cite{wales03}.  The rate constants $k_{uv}$ were calculated according to transition
state theory.  All numerical computations were performed using our public domain
software packages GMIN, OPTIM, and PATHSAMPLE. %  \cite{gmin,optim,pathsample}.  

The examples considered here correspond to relatively sparse networks of
between 2000 and 100000 nodes.  Typically the number of edges was several times
the number of nodes.  Hence we compared NGT with the C-language sparse linear
algebra package UMFPACK \cite{umfpack}, which employs sparse LU factorization.  
UMFPACK is contained in the python
scientific computing package SciPy \cite{scipy}.  
% UMFPACK is short for
% `unsymmetric multifrontal sparse LU factorization package' and is also used in
% Matlab as the default engine for sparse unsymmetric LU decomposition.  
We tried several other methods for solving the linear equations, including
SuperLU \cite{superlu}, another sparse LU decomposition package; conjugate
gradient iteration \cite{scipy}; and, after symmetrizing $R$, sparse
Cholesky decomposition using CHOLMOD \cite{cholmod}.  All of these
methods exhibited similar or poorer performance than UMFPACK.

We report here only the results for the MFPTs.
Computing committors for NGT uses exactly the same procedure.  Calculating
committors with linear algebra requires solving a different system of linear
equations, but the performance results are very similar to those for the mean
first passage times.

The results for computing the MFPTs between two groups of nodes
are shown in figures \ref{fig:lj38_times} and \ref{fig:lj75_times} for the LJ$_{38}$ and LJ$_{75}$ clusters,
and figure \ref{fig:beta3s_times} for the three-stranded $\beta$ sheet peptide.
When the procedures agree, the sparse linear solver is
about 1.5 times faster than NGT for LJ$_{38}$, and about an order of magnitude faster for LJ$_{75}$.
However, the linear algebra approaches
often fail, returning unphysical results, such as 
negative mean first passage times.

The linear algebra solvers fail more often for larger systems, and rarely work for the lower temperatures
that are the main focus of interest for rare event dynamics.
For low temperatures, 
the largest and smallest relaxation times can differ by many orders of magnitude.  This ill-conditioning leads to the possibility of 
large errors arising from numerical imprecision.

The special property of the rate matrix $\sum_v R_{uv} = 0$ means that precision issues
are a problem from the beginning.  This property is reflected
in the transition probabilities, which conserve the total probability.  
The problem can be understood by the fact that a floating point number
cannot precisely represent values arbitrarily close
to zero or arbitrarily close to one.  The NGT method was specifically designed
to solve these problems.  At every step in the graph renormalization, the transition probabilities
at each node $u$ satisfy $\sum_v P_{uv} = 1$.  This condition means that when computing $1-P_{uu}$ we can either use 
$1-P_{uu}$ directly or indirectly via $\sum_{v \ne u} P_{uv}$.  
In practice we use the former definition unless $P_{uu} > 0.99$.  We believe this procedure accounts for the fact that 
the linear algebra method fails regularly, while NGT always produces a sensible result.
It is possible that a preconditioning procedure could be derived that increases
the stability of the linear algebra method, but 
we have not yet found a method that improves the
present results.

%\subsection{Conclusion}

In summary,
we have compared the performance of the NGT algorithm for computing mean first
passage times and committor probabilities with sparse linear algebra packages.
We have shown that the linear algebra packages can be somewhat faster, but
frequently fail at the lower temperatures of interest.  We
believe that this result is due to problems with numerical precision, which
occur when the ratio of the largest relaxation time to the smallest is large.
The NGT algorithm avoids these numerical problems by taking advantage of the
physical structure of the problem to precisely represent important
probabilities that are arbitrarily close to zero or unity.

Systems that exhibit multifunnel energy landscapes \cite{doyemw99},
with competing morphologies separated by high barriers, exhibit interesting
properties. Low temperature heat capacity peaks correspond to broken ergodicity,
and multiple relaxation time scales reflect rare event dynamics \cite{Wales10a}.
Such landscapes present significant challenges for global optimisation and sampling.
Recent developments for analysing thermodynamics 
\cite{NeirottiCFD00,CalvoNFD00,MandelshtamFC06,SharapovMM07,SharapovM07,Calvo10,SehgalMF14}
and kinetics \cite{PiccianiAKT11}
will enable us to validate the approximations that make computational potential energy landscape
approaches, such as basin-sampling \cite{BogdanWC06,Wales2013} and discrete path sampling \cite{wales02,wales04,Wales06}, so efficient.
The present work provides another key piece of information, confirming the 
accuracy of the NGT procedure for extracting rates from kinetic transition networks,
and the efficiency of the method for treating the dynamics of multi-funnel landscapes.

\begin{acknowledgements}
  We gratefully acknowledge Eric Vanden-Eijnden for helpful discussions.  We
  also thank the EPSRC and European Research Council for support.
\end{acknowledgements}

%\bibliography{../../bib/wales,../../bib/cluster.thermo,../../bib/books,../../bib/rareevent,../../bib/landscapes,../../bib/rates,../../bib/kmc}
%\bibliography{ngt_benchmark}

\begin{thebibliography}{10}

\bibitem{NoeF08}
F.~No\'e and S.~Fischer,
\newblock Curr. Op. Struct. Biol. {\bf 18}, 154 (2008).

\bibitem{pradag09}
D.~Prada-Gracia, J.~G\'omez-Gardenes, P.~Echenique, and F.~Fernando,
\newblock PLoS Comput. Biol. {\bf 5}, 1 (2009).

\bibitem{masoliver.2006}
J.~Masoliver, M.~Montero, J.~Perello, and G.~H. Weiss,
\newblock J. Economic Behavior and Organization {\bf 61}, 577 (2006).

\bibitem{acemoglu.2013}
D.~Acemoglu, G.~Como, F.~Fagnani, and A.~Ozdaglar,
\newblock Math. Oper. Res. {\bf 38}, 1 (2013).

\bibitem{NoeSVRW09}
F.~Noe, C.~Schutte, E.~Vanden-Eijnden, L.~Reich, and T.~R. Weikl,
\newblock Proc. Nat. Acad. Sci. USA {\bf 106}, 19011 (2009).

\bibitem{wales02}
D.~J. Wales,
\newblock Mol. Phys. {\bf 100}, 3285 (2002).

\bibitem{wales04}
D.~J. Wales,
\newblock Mol. Phys. {\bf 102}, 891 (2004).

\bibitem{HenkelmanJ01}
G.~Henkelman and H.~J\'onsson,
\newblock J. Chem. Phys. {\bf 115}, 9657 (2001).

\bibitem{glowacki.2012}
D.~R. Glowacki, C.-H. Liang, C.~Morley, M.~J. Pilling, and S.~H. Robertson,
\newblock J. Phys. Chem. A {\bf 116}, 9545 (2012).

\bibitem{Wales09}
D.~J. Wales,
\newblock J. Chem. Phys. {\bf 130}, 204111 (2009).

\bibitem{Wales10a}
D.~J. Wales,
\newblock Curr.~Op.~Struct.~Biol. {\bf 20}, 3 (2010).

\bibitem{norris.1997}
J.~R. Norris,
\newblock {\em Markov Chains},
\newblock Cambridge University Press, 1997.

\bibitem{metznersv09}
P.~Metzner, C.~Sch\"utte, and E.~Vanden-Eijnden,
\newblock Multiscale Model. Simul. {\bf 7}, 1192 (2009).

\bibitem{trygubenkoW06a}
S.~A. Trygubenko and D.~J. Wales,
\newblock Mol. Phys. {\bf 104}, 1497 (2006).

\bibitem{trygubenkow06b}
S.~A. Trygubenko and D.~J. Wales,
\newblock J. Chem. Phys. {\bf 124}, 234110 (2006).

\bibitem{landau.2005}
D.~Landau and K.~Binder,
\newblock {\em A Guide to Monte Carlo Simulations in Statistical Physics},
\newblock Cambridge University Press, New York, NY, USA, 2005.

\bibitem{doyemw99}
J.~P.~K. Doye, M.~A. Miller, and D.~J. Wales,
\newblock J. Chem. Phys. {\bf 110}, 6896 (1999).

\bibitem{carrw08}
J.~M. Carr and D.~J. Wales,
\newblock J. Phys. Chem. B {\bf 112}, 8760 (2008).

\bibitem{wales03}
D.~J. Wales,
\newblock {\em Energy Landscapes},
\newblock Cambridge University Press, Cambridge, 2003.

\bibitem{umfpack}
T.~A. Davis,
\newblock ACM Trans. Math. Software {\bf 30}, 196 (2004).

\bibitem{scipy}
E.~Jones et~al.,
\newblock {SciPy}: Open source scientific tools for {Python}, 2001--.

\bibitem{superlu}
X.~S. Li,
\newblock ACM Trans. Math. Software {\bf 31}, 302 (2005).

\bibitem{cholmod}
Y.~Chen, T.~A. Davis, W.~W. Hager, and S.~Rajamanickam,
\newblock ACM Trans. Math. Software {\bf 35} (2009).

\bibitem{NeirottiCFD00}
J.~P. Neirotti, F.~Calvo, D.~L. Freeman, and J.~D. Doll,
\newblock J. Chem. Phys. {\bf 112}, 10340 (2000).

\bibitem{CalvoNFD00}
F.~Calvo, J.~P. Neirotti, D.~L. Freeman, and J.~D. Doll,
\newblock J. Chem. Phys. {\bf 112}, 10350 (2000).

\bibitem{MandelshtamFC06}
V.~A. Mandelshtam, P.~A. Frantsuzov, and F.~Calvo,
\newblock J. Phys. Chem. A {\bf 110}, 5326 (2006).

\bibitem{SharapovMM07}
V.~A. Sharapov, D.~Meluzzi, and V.~A. Mandelshtam,
\newblock Phys. Rev. Lett. {\bf 98}, 105701 (2007).

\bibitem{SharapovM07}
V.~A. Sharapov and V.~A. Mandelshtam,
\newblock J. Phys. Chem. A {\bf 111}, 10284 (2007).

\bibitem{Calvo10}
F.~Calvo,
\newblock Phys. Rev. E {\bf 82}, 046703 (2010).

\bibitem{SehgalMF14}
R.~M. Sehgal, D.~Maroudas, and D.~M. Ford,
\newblock J. Chem. Phys. {\bf 140},  (2014).

\bibitem{PiccianiAKT11}
M.~Picciani, M.~Athenes, J.~Kurchan, and J.~Tailleur,
\newblock J. Chem. Phys. {\bf 135}, 034108 (2011).

\bibitem{BogdanWC06}
T.~V. Bogdan, D.~J. Wales, and F.~Calvo,
\newblock J. Chem. Phys. {\bf 124}, 044102 (2006).

\bibitem{Wales2013}
D.~J. Wales,
\newblock Chem. Phys. Lett. {\bf 584}, 1  (2013).

\bibitem{Wales06}
D.~J. Wales,
\newblock Int. Rev. Phys. Chem. {\bf 25}, 237 (2006).

\end{thebibliography}

\end{document}